\documentclass[reprint,amsmath,amssymb,aps,]{revtex4-1}
\usepackage{xcolor}
\usepackage{siunitx}
\usepackage{graphicx}
\usepackage{dcolumn}
\usepackage{bm}

\begin{document}

\preprint{APS/123-QED}

\title{Ferromagnetic Quantum Critical Point in CePd$_2$P$_2$ with Pd $\rightarrow$ Ni Substitution}

\author{Y. Lai$^{1,2}$, S. E. Bone$^{3}$, S. Minasian$^{4}$, M. G. Ferrier$^{3}$, J. Lezama-Pacheco$^{5}$, V. Mocko$^{3}$, A. S. Ditter$^{3,6}$, S. A. Kozimor$^{3}$, G. T. Seidler$^{6}$, W. L. Nelson$^{1,2}$, Y.-C. Chiu$^{1,2}$,  K. Huang$^{1}$, W. Potter$^{7}$, D. Graf$^{1}$, T. E. Albrecht-Schmitt$^{7}$, R. E. Baumbach$^{1,2}$}

\email{baumbach@magnet.fsu.edu}
\affiliation{$^1$National High Magnetic Field Laboratory, Florida State University, Tallahassee, FL 32310, USA}
\affiliation{$^2$Department of Physics, Florida State University, Tallahassee, FL 32306, USA}
\affiliation{$^3$Los Alamos National Laboratory, Los Alamos, NM 87544} 
\affiliation{$^4$Lawrence Berkeley National Laboratory, Berkeley, CA 94720}
\affiliation{$^5$Stanford University, Stanford, CA 94305}
\affiliation{$^6$University of Washington, Seattle WA 98195}
\affiliation{$^7$Department of Chemistry and Biochemistry, Florida State University,Tallahassee, FL 32306, USA}

\date{\today}
\begin{abstract}
An investigation of the structural, thermodynamic, and electronic transport properties of the isoelectronic chemical substitution series Ce(Pd$_{1-x}$Ni$_x$)$_2$P$_2$ is reported, where a possible ferromagnetic quantum critical point is uncovered in the temperature - concentration ($T-x$) phase diagram. This behavior results from the simultaneous contraction of the unit cell volume, which tunes the relative strengths of the Kondo and RKKY interactions, and the introduction of disorder through alloying. Near the critical region at $x_{\rm{cr}}$ $\approx$ 0.7, the rate of contraction of the unit cell volume strengthens, indicating that the cerium $f$-valence crosses over from trivalent to a non-integer value. Consistent with this picture, x-ray absorption spectroscopy measurements reveal that while CePd$_2$P$_2$ has a purely trivalent cerium $f$-state, CeNi$_2$P$_2$ has a small ($<$ 10 \%) tetravalent contribution. In a broad region around $x_{\rm{cr}}$, there is a breakdown of Fermi liquid temperature dependences, signaling the influence of quantum critical fluctuations and disorder effects. Measurements of clean CePd$_2$P$_2$ furthermore show that applied pressure has a similar initial effect to alloying on the ferromagnetic order. From these results, CePd$_2$P$_2$ emerges as a keystone system to test theories such as the Belitz-Kirkpatrick-Vojta model for ferromagnetic quantum criticality, where distinct behaviors are expected in the dirty and clean limits.

\begin{description}

\item[PACS numbers]
PACS

\end{description}
\end{abstract}

\pacs{Valid PACS appear here}

\maketitle

\section{\label{sec:level1}Introduction}
Heavy fermion $f$-electron intermetallics continue to attract interest because many of them exhibit complex phase diagrams with diverse phenomena including nematic electronic states, charge and spin instabilities, and unconventional superconductivity.~\cite{Stewart01,Rosch07,Gegenwart08,Pfleiderer09,brando16} In many cases this is related to a competition between the RKKY and Kondo interactions, which mediate magnetism and compensate localized spins, respectively.~\cite{Doniach_77,kondo,ruderman,kasuya,yosida} The fine balance between these interactions can cause a magnetic ordering temperature to be continuously suppressed towards zero temperature at a quantum critical point (QCP). As this occurs novel behaviors often emerge including the breakdown of Fermi liquid behavior and the emergence of superconductivity.~\cite{Pfleiderer09,mathur98,thompson12} This has led to a viewpoint that quantum critical fluctuations of an order parameter are key for producing novel phenomena, and this phenomenology even spans diverse families of materials that are distinct from $f$-electron intermetallics: e.g., cuprate and iron-based superconductors,~\cite{ramshaw15,paglione10} organic superconductors,~\cite{leyraud09} charge density wave systems,~\cite{gruner17} and others.

There nonetheless remain many open questions, including how the specific type of magnetism being suppressed influences a quantum critical region. The earliest theories of quantum phase transitions focused on ferromagnetism,~\cite{stoner} and the seminal work of Hertz and Millis predicted that a ferromagnetic phase transition would remain continuous to zero temperature.~\cite{hertz,millis} More recent work by Belitz-Kirkpatrick-Vojta (BKV) demonstrates instead that for clean materials in two and three dimensions a zero temperature transition from ferromagnetism to paramagnetism is discontinuous.~\cite{brando16, Belitz12, belitz99,belitz05,belitz15} The first order change at the phase boundary prevents diverging fluctuations of the magnetic order parameter. This is in contrast to what is seen near second order antiferromagnetic QCPs where the order parameter diverges, and may interfere with phenomena such as unconventional superconductivity. The BKV theory also predicts that there is a tricritical point that separates a high temperature line of second order phase transitions from a low temperature line of first order phase transitions, where the application of a magnetic field produces wing-like second order phase boundaries that intercept zero temperature. In disordered systems the tricritical point is pushed below zero temperature and the second order phase boundary extends to zero temperature. This has spurred interest in ferromagnetic quantum criticality in disordered metals, where an intriguing possibility is that they might host anomalous metallic states and even unconventional superconductivity.~\cite{belitz99, belitz05, belitz15, brando16,huang13, Nakatsuji08} It is noteworthy that while there are U- and Yb- based ferromagnetic superconductors~\cite{saxena2001,huy,aoki2001,akazawa_2004,ner11}, there are no cerium-based analogues despite some electronic similarities between the associated 5$f$ and 4$f$ states.

CePd$_2$P$_2$ was recently reported to be a correlated electron ferromagnet crystallizing in the well-known ThCr$_2$Si$_2$-type structure~\cite{tran14,ikeda15}, while its isoelectronic volume compressed analogue CeNi$_2$P$_2$ exhibits a nonmagnetic ground state.~\cite{jeitschko,chen,Dra} This suggests that the Pd $\rightarrow$ Ni alloy series could host a ferromagnetic QCP. We synthesized single crystal specimens of Ce(Pd$_{1-x}$Ni$_x$)$_2$P$_2$ for 0 $<$ $x$ $<$ 1, where the contracting unit cell volume applies a chemical pressure. X-ray diffraction and magnetic susceptibility measurements show that the cerium ions remain nearly trivalent up to $x$ $\approx$ 0.66, where the rate of unit cell volume contraction increases, signaling a change in the $f$-electron valence. X-ray absorption spectroscopy measurements for CePd$_2$P$_2$ and CeNi$_2$P$_2$ reinforce this view by revealing trivalent and trivalent with a small fraction of tetravalent $f$-electron character, respectively. Features associated with the ferromagnetic ordering are evident for 0 $<$ $x$ $\lesssim$ 0.69 in the magnetic susceptibility, heat capacity, and electrical resistivity, where the ordering temperature is continuously suppressed towards zero at an extrapolated critical value of $x_{\rm{cr}}$ $\approx$ 0.7. In the critical region there is chemical disorder which allows the phase transition to remain second order, even as it approaches zero temperature. This results in a putative ferromagnetic QCP, around which there are indications for a breakdown of Fermi liquid behavior: in particular, the heat capacity divided by temperature $C/T$ diverges nearly logarithmically with decreasing $T$. There is also evidence that the disorder contributes to the unusual temperature dependences by producing a quantum Griffiths phase that extends over a broad $x$-range.~\cite{Griffiths1969,vojta,sereni07, Westerkamp09} We furthermore find that for clean CePd$_2$P$_2$, applied pressure initially suppresses the ferromagnetism in a manner similar to that of Pd $\rightarrow$ Ni substitution. Therefore, this system offers the opportunity to study behavior at a disordered ferromagnetic quantum critical point and eventually to compare to the ordered analogue, as is needed to test the BKV theory and to ultimately design new QCP materials.

\section{\label{sec:level1}Experimental Methods}
Single crystals of Ce(Pd$_{1-x}$Ni$_x$)$_2$P$_2$ were grown from elements with purities \(>\) 99.9\% in a molten flux of Pd, Ni and P. The reaction ampoules were prepared by loading the elements in the ratio Ce:Pd:Ni:P $;$ 1:11(1-$x$):11$x$:8 into a 2 mL alumina crucible for each of the nominal Ni concentrations. The crucibles were sealed under vacuum in quartz ampoules and heated to $300\,^{\circ}\mathrm{C}$ at a rate of $50\,^{\circ}\mathrm{C}$/hour, held at $300\,^{\circ}\mathrm{C}$ for 6 hours, heated to $500\,^{\circ}\mathrm{C}$ at a rate of $50\,^{\circ}\mathrm{C}$/hour, held at $500\,^{\circ}\mathrm{C}$ for 6 hours, heated to $1180\,^{\circ}\mathrm{C}$ at a rate of $50\,^{\circ}\mathrm{C}$/hour, kept at $1180\,^{\circ}\mathrm{C}$ for 3 hours, and then cooled at a rate of $2\,^{\circ}\mathrm{C}$/hour to $1000\,^{\circ}\mathrm{C}$. At this temperature, the remaining flux was separated from the crystals by centrifuging. Single-crystal platelets with typical dimensions of several millimeters on a side and several millimeters in thickness were collected.

The crystal structure and chemical composition were verified by powder x-ray-diffraction (XRD) and energy dispersive spectrometer (EDS) analysis. EDS results are shown in Fig.~\ref{xrd}a, where the measured concentration $x_{\rm{act}}$ is compared to the nominal concentration $x_{\rm{nom}}$. Throughout the rest of the manuscript we use $x_{\rm{act}}$ unless otherwise specified. Magnetization $M$($T,H$) measurements were carried out for single crystals at temperatures $T$ $=$ 1.8 $-$ 300 K under an applied magnetic field of $H$ $=$ 5 kOe for $H$ applied both parallel ($\parallel$) and perpendicular ($\perp$) to the $c$ axis using a Quantum Design VSM Magnetic Property Measurement System. The AC magnetic susceptibility $\chi$'($T$) for selected concentrations was also measured using the same apparatus. Electrical resistivity $\rho$ measurements for temperatures $T$ $=$ 0.5 $-$ 300 K were performed in a four-wire configuration and the heat capacity $C$ was measured for $T$ $=$ 0.39 $-$ 20 K using a Quantum Design Physical Property Measurement System. $\rho$($T$) measurements under applied pressure were performed using a piston cylinder pressure cell with Daphne 7474 oil as the pressure transmitting medium. The pressure is determined by the shift in ruby flourescence peaks and are the values determined below $T$ $=$ 10 K. These measurements were performed at the National High Magnetic Field Laboratory DC field User facility using standard He3 cryostats.

Samples were analyzed using Ce L$_3$-edge X-ray absorption spectroscopy (XANES) at the Stanford Synchrotron Radiation Lightsource (SSRL) on beam line 11-2. Single crystals of the compounds were ground and diluted with boron nitride and painted onto 0.5 mil Kapton tape, the tape was attached to an aluminum sample plate and loaded into a liquid helium cryostat. A single energy was selected using a liquid-N2-cooled double-crystal monochromator utilizing Si(220) ($\phi$ = 0) crystals. The crystals were detuned by 70\% at 6100 eV to remove higher order harmonics. Spectra were measured in fluorescence mode using a Lytle detector equipped with a Ti filter (3 absorption lengths) at two different temperatures, 85 K and 10 K. A Cr-calibration foil was placed downstream of the sample and spectra were calibrated to the first reflection point of Cr(5989.0 eV). Using the Athena11 software package, L$_3$-edge spectral were background subtracted and normalized at $E_0$ (5723 eV). A deconvoluted model for the Ce L$_3$-edge XANES data was obtained using a modified version of EDG-FIT~\cite{nolas} in IGOR 6.0. Using this least-squares algorithm spectra were modeled with a minimum number of pseudo-Voight functions (50:50 Lorentzian:Gaussian) and a 1:1 ratio of arctangent and error function. The areas under the pre-edge peaks (hereafter defined as the intensity) were equal to the FWHM x peak height.

Single energy images, elemental maps, and Ce M$_{5,4}$-edge x-ray absorption spectra (XANES)~\cite{ev07} were acquired using the scanning transmission x-ray microscope (STXM) instrument at the spectromicroscopy beamline 10ID-1 at the Canadian Light Source (CLS) according to data acquisition methodology described previously.~\cite{min17,man18, ha18}

\section{\label{sec:level1}Results}
\begin{figure}[!tht]
	\begin{center}
		\includegraphics[width=1\linewidth]{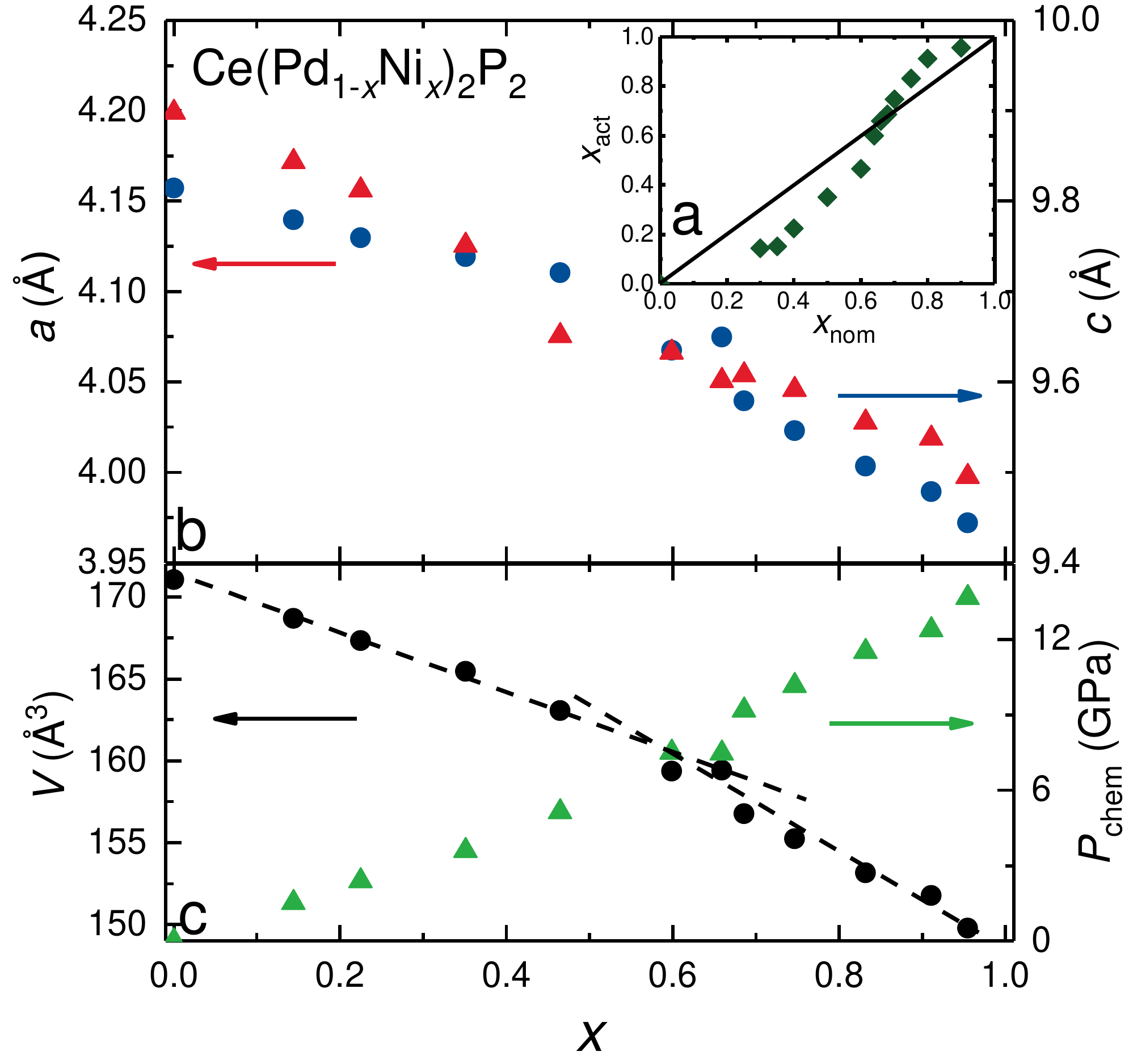}
		\caption{ (a) Comparison between the measured phosphorous concentration $x_{\rm{act}}$ and the nominal concentration $x_{\rm{nom}}$, where $x_{\rm{act}}$ was determined using energy dispersive spectrometer analysis. Throughout the rest of the manuscript we use $x_{\rm{act}}$ $=$ $x$ unless otherwise specified. (b) The lattice constants, $a(x)$ (left axis) and $c(x)$ (right axis).  (c) the unit cell volume $V(x)$ (left axis) and chemical pressure $P_{\rm{ch}}$(right axis), calculated using the Birch-Murnaghan equation as described in the text with the bulk modulus $B_{0}$ $=$ 110 GPa. 
		}
		\label{xrd}
	\end{center}
\end{figure}

Powder X-ray diffraction measurements show that the ThCr$_2$Si$_2$-type structure persists across entire the Pd $\rightarrow$ Ni substitution series, while the tetragonal lattice constants ($a$ and $c$) and the unit cell volume ($V$) decrease with increasing $x$ (Fig.~\ref{xrd}). Up to $x$ $\approx$ 0.66 the trends are consistent with Vegard's law, where the linear lattice contraction is due to the smaller size of Ni by comparison to Pd. This suggests that over this $x$-range the room temperature Ce valence remains constant. For $x$ $\gtrsim$ 0.66 the unit cell volume continues to decrease linearly, but with a larger slope, signaling a change in the cerium valence. The volume contraction results in a chemical pressure which is estimated to be near $P_{\rm{ch}}$ $=$ 7.5 GPa for $x$ $=$ 0.66 and $P_{\rm{ch}}$ $=$ 13.6 GPa at $x$= 0.96. These values are calculated using the Birch-Murnaghan equation $P_{\rm{ch}}$ $=$ $B_{\rm{0}}$$\Delta$$V(x)/V(0)$, where the value of the bulk modulus for CeCu$_2$Si$_2$ ($B_{\rm{0}}$ $=$ 110 GPa) is used.~\cite{spain86}

\begin{figure}[!tht]
    \begin{center}
        \includegraphics[width=1\linewidth]{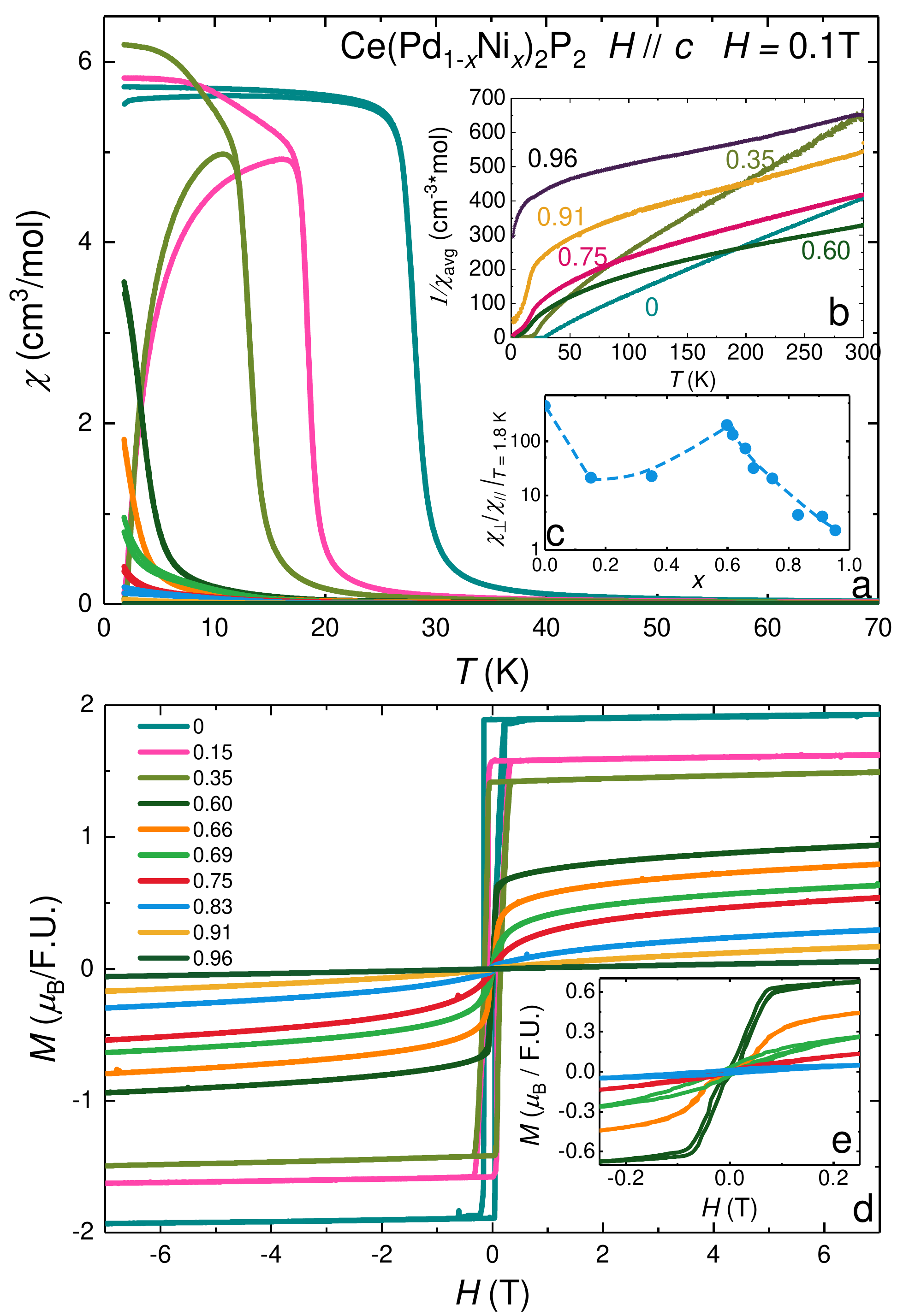}
        \caption{(a) Magnetic susceptibility $\chi$ $=$ $M/H$ vs temperature $T$ for $H$ $=$ 0.1 T applied parallel $\parallel$ to the $c$-axis for Ce(Pd$_{1-x}$Ni$_x$)$_2$P$_2$. (b) The inverse of Magnetic susceptibility $\chi_{\rm{avg}}$$^{-1}$($T$) for representative curves, where $\chi_{\rm{avg}}$ = (2$\chi_{ab}$ + $\chi_c$)/3. The dotted lines are Curie-Weiss fits to the data for 150 K $<$ $T$ $<$ 300 K. (c) The magnetic anisotropy $\chi_{\perp}$/$\chi_{\parallel}$ vs $x$ at $T$ $=$ 2 K. (d) Magnetization field dependence for different $x$-concentrations measured at 1.8K. (e) Zoom in low field of $M$ vs $H$ for different $x$-concentrations measured at 1.8K.
        }
        \label{chi}
    \end{center}
\end{figure}

The magnetic susceptibility $\chi$ $=$ $M/H$ vs temperature and magnetization $M$ vs $H$ for $H$ $\parallel$ $c$ data are shown in Fig.~\ref{chi}. As previously reported for polycrystalline specimens,~\cite{tran14} ferromagnetic ordering appears in $\chi(T)$ for $x$ $=$ 0 as a sharp increase at $T_{\rm{C}}$ $=$ 28.5 K, which we define as the peak in $\partial$$\chi$/$\partial$$T$ (not shown). For $T$ $\leq$ $T_{\rm{C}}$, $M(H)$ rapidly saturates towards $M_{\rm{sat}}$ $=$ 1.93 $\mu_{\rm{B}}$. The evolution of the ferromagnetic order with $x$ is determined using these quantities, where $T_{\rm{C}}$ decreases linearly and is extrapolated to intercept zero temperature near $x_{\rm{cr}}$ $\approx$ 0.7. The persistence of the ferromagnetism into the critical $x$-region is seen in the $M(H)$ curves, which remain hysteretic even as $M_{\rm{sat}}$ is smoothly suppressed (Figs.~\ref{chi}d and e). For $x$ $\gtrsim$ $x_{\rm{cr}}$, the magnitudes of $\chi(T)$ and $M(H)$ continue to decrease and become similar to that of paramagnetic CeNi$_2$P$_2$ as $x$ approaches 1.~\cite{chen} 

The high temperature Curie-Weiss behavior provides further insight into the $x$-evolution of the $f$-electron state, magnetocrystalline anisotropy, and strength of the Kondo hybridization. There is a strong evolution in the  magnetic anisotropy $\chi_{\perp}$/$\chi_{\parallel}$ between the $c$ and $a$$b$ directions: while it decreases by roughly a factor of 10 to approach $\chi_{\perp}$/$\chi_{\parallel}$ $\approx$ 20 near $x$ $\approx$ 0.35, it recovers to nearly the $x$ $=$ 0 value at $x$ $\approx$ 0.6 and finally decreases to become isotropic at $x$ $=$ 1 (Fig.~\ref{chi}c). In order to analyze this data using Curie-Weiss fits, we calculate the average susceptibility, defined as $\chi_{\rm{avg}}$ = (2$\chi_{ab}$ + $\chi_c$)/3 (Fig.~\ref{chi}b). For $x$ = 0, $\chi_{\rm{avg}}$ is consistent with earlier results for polycrystalline specimens~\cite{tran14}, yielding an effective magnetic moment $\mu_{\rm{eff}}$ = 2.4 $\mu_{\rm{B}}$ (trivalent cerium) and a Curie-Weiss temperature $\theta$ = 2 K. Pd $\rightarrow$ Ni substitution causes $\theta$ to increase to a value near -193 K at $x$ = 0.66, and  afterwards to even larger negative values. This is a common feature  in Ce-based materials with strong hybridization between the $f$- and conduction electrons and indicates that the Kondo interaction strengthens with increasing $x$.~\cite{sereni07}

\begin{figure}[!tht]
	\begin{center}
		\includegraphics[width=1\linewidth]{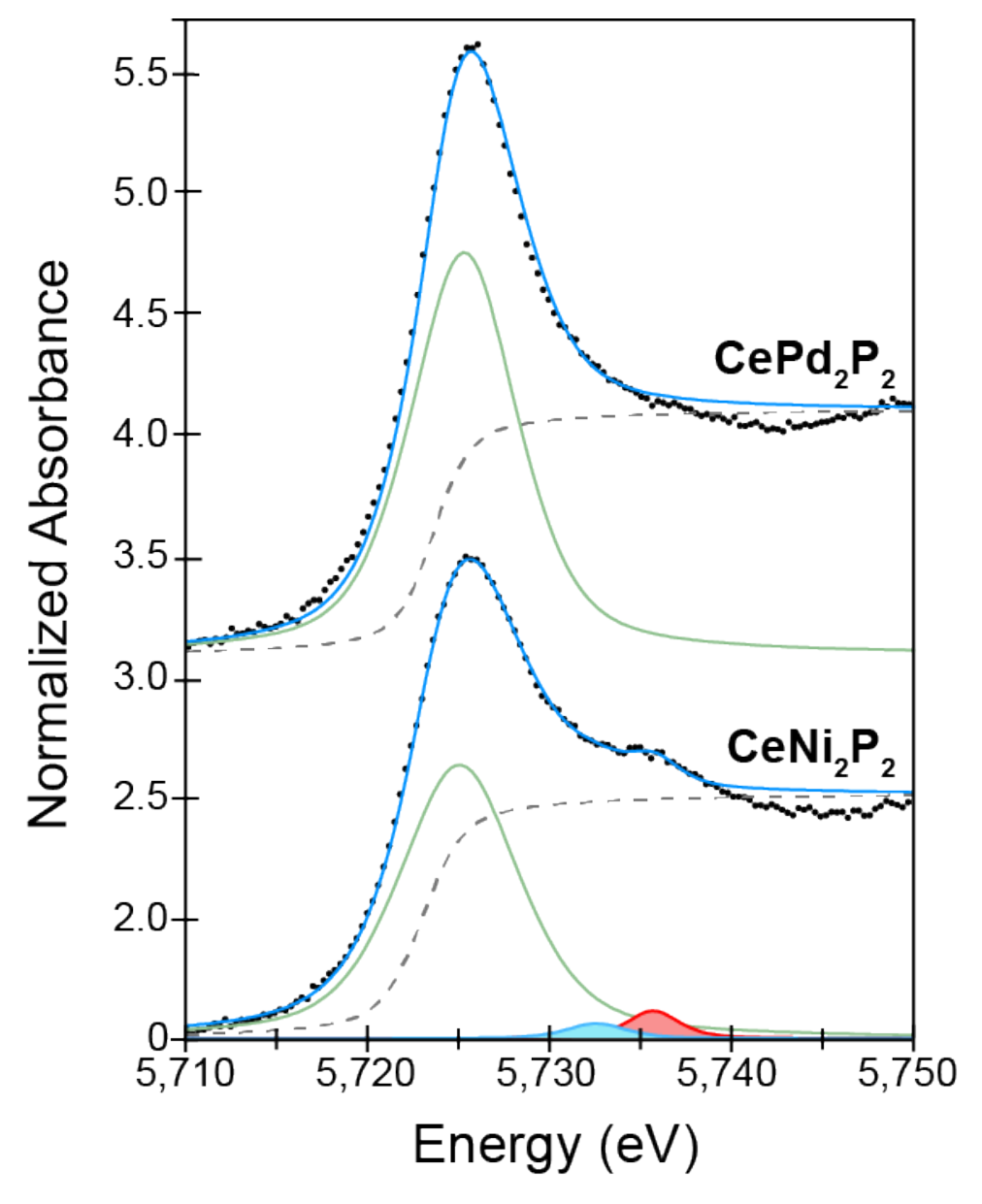}
		\caption{The experimental data (dots) and the curve fitted model (blue trace) for the Ce L$_3$-edge X-ray absorption spectra of CePd$_2$P$_2$ (top) and CeNi$_2$P$_2$ (bottom). The pre-edge pseudo-Voigt functions (green, blue, and red traces) used to generate the model and the step function (grey dashed lines) are shown. 
		}
		\label{xas}
	\end{center}
\end{figure}

To further evaluate the effect of Pd $\rightarrow$ Ni substitution  on the Ce $f$-electron state, Ce L$_3$-edge X-ray absorption spectra (XANES) were obtained from  single crystals of CeNi$_2$P$_2$ and CePd$_2$P$_2$, the end members  of the series. Ce$^{3+}$ usually show a single absorption peak at ca. 5723-5725 eV, whereas Ce$^{4+}$ exhibits a ``double-white line" feature with maxima at ca. 5724-5728 and 5736-5739 eV. As shown in Fig.~\ref{xas}, the Ce L$_3$-edge spectrum from CePd$_2$P$_2$ is typical of Ce$^{3+}$, containing a single pronounced absorption peak with a maximum at 5725.2 eV (FWHM = 6.60 eV). Although substituting Ni for Pd has essentially no impact on the main absorption peak energy (maximum = 5725.1 eV), subtle spectral changes emerges; most notably, the main absorption peak broadens by 1.24 eV (FWHM = 7.84 eV) and a minor post-edge feature emerges near 5735 eV. Additionally, spectra obtained from CeNi$_2$P$_2$ and CePd$_2$P$_2$ are unchanged as a function of temperature between 85 and 10 K.

To characterize the origin of these changes, the Ce L$_3$-edge XANES spectra are modeled using a least-squares algorithm between 5705 and 5741 eV. Although the entire fitted region for CePd$_2$P$_2$ is easily modeled by the combination of a single peak at 5725.2 eV and a step-function at 5724.4 eV, an analogous model for CeNi$_2$P$_2$ does not adequately represent the data. In particular, there is substantial misfit associated with the post-edge feature that is absent in the CePd$_2$P$_2$ spectrum. Hence, three peaks and a step-function are needed to model the CeNi$_2$P$_2$ spectrum. The main absorption peak is at 5725.1 eV, the step function at 5723.0 eV, and two post-edge peaks are fit at 5732.6 and 5735.7 eV. We believe it is of no coincidence that the highest-energy post-edge  peak (at 5735.7 eV) occurs  at  a similar energy to the higher energy peak of the ``double-white line" feature typically observed for Ce$^{4+}$ (between 5736 and 5739 eV).~\cite{
bianconi1987,kaindl1988,sham2005,walter2009,loble2015,bogart2015,cary2016,kratsova2016,antonio2017,toscani2016} Hence, these data are interpreted as indicating that CeNi$_2$P$_2$ contains a  mixture  of  Ce$^{3+}$  and  Ce$^{4+}$. Comparison of the intensities (FWHM x peak height) of the main absorption peak (intensity = 8.8) with the small post-edge peaks (intensities = 0.2 and 0.4) suggests that CeNi$_2$P$_2$ contains on the order of 7(1)\% Ce.

\begin{figure}[!tht]
	\begin{center}
		\includegraphics[width=1\linewidth]{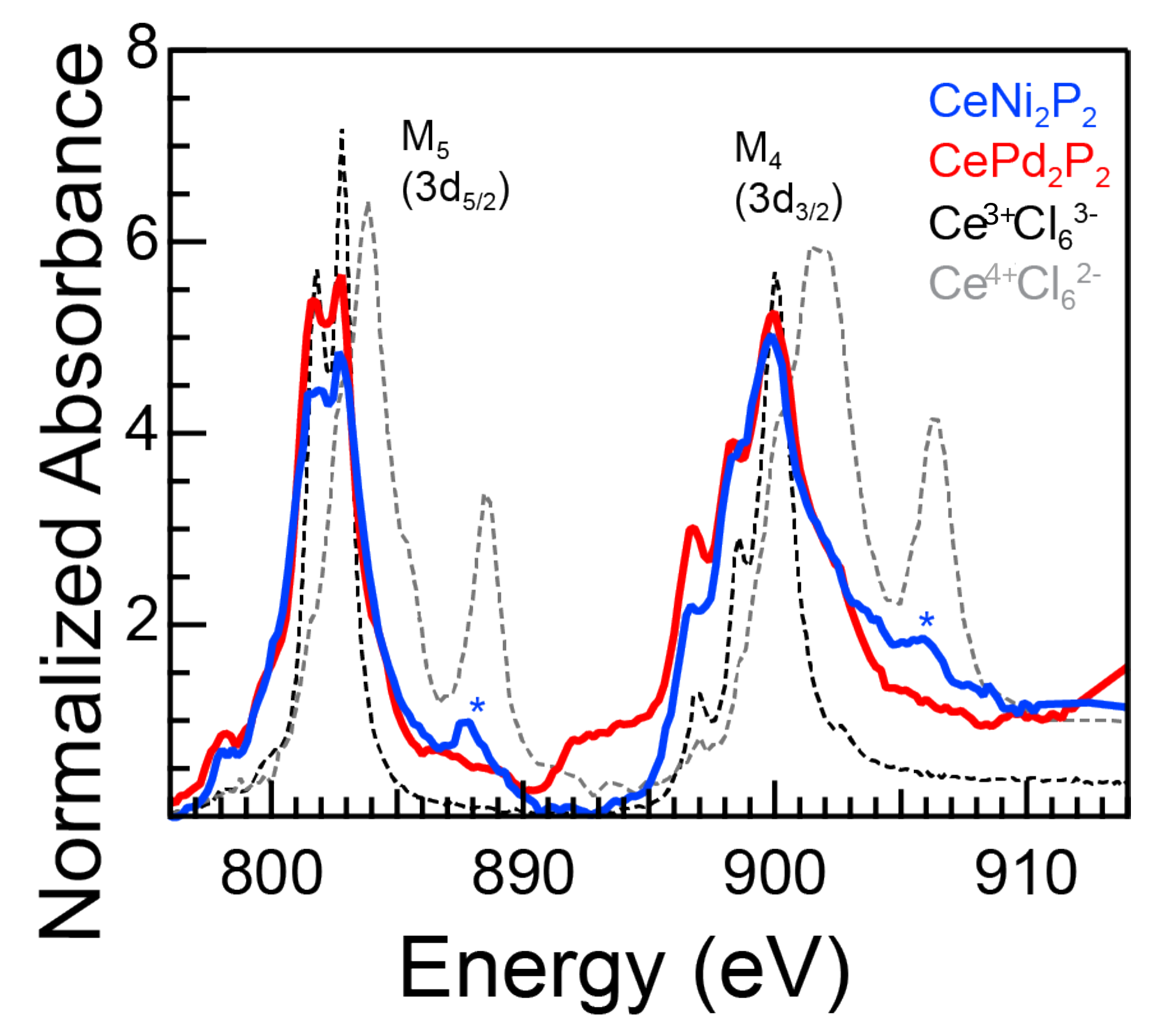}
		\caption{Cerium M$_{5,4}$-edge XANES spectra for CeNi$_2$P$_2$ (blue trace) and CePd$_2$P$_2$ (red trace) along with Ce$^{3+}$ and Ce$^{4+}$ reference compounds (dashed black and gray traces). The [Ph$_4$P]$_3$CeCl$_6$ and [Et$_4$N]$_2$CeCl$_6$ data are adapted with permission from ref.~\cite{loble2015}. 
		}
		\label{XANES}
	\end{center}
\end{figure}

XANES data at the M$_{5,4}$-edge were obtained to corroborate the L$_3$-edge measurements. The spectroscopic approach can be advantageous for probing 4$f$ orbital occupation and mixing, especially for systems with mixed valence or multiconfigurational ground states.~\cite{bianconi1987,dex87,kaindl1988,ani11, ani12,hu00,kaindl84,le85,ra92} The background subtracted and normalized M$_{5,4}$-edge spectra for CeNi$_2$P$_2$ and CePd$_2$P$_2$ are provided in Fig.~\ref{XANES}. The M$_{5,4}$-edge spectra are split into low energy M$_5$ (3$d_{5/2}$) and high energy M$_4$ (3$d_{3/2}$) edges due to spin-orbit coupling with the 3$d$ core hole. For both CeNi$_2$P$_2$ and CePd$_2$P$_2$ as well as the CeCl$_6$$^{3-}$ reference, the M$_{5,4}$-edge exhibits a characteristic ``sawtooth" pattern with fine structure that closely resembles expectations from theory for a 3$d^9$4$f^2$ final state and an isolated Ce$^{3+}$ ion.~\cite{le85} For the Ce$^{4+}$ reference, CeCl$_6$$^{2-}$, both the M$_5$- and M$_4$-edges are split into intense main peaks and additional satellite features about 5 eV higher in energy.~\cite{le85, am16, hu97, jean98, dong03} Upon close inspection, weak satellite features are also present 887.3 and 905.4 eV in the Ce M$_{5,4}$-edge spectrum for CeNi$_2$P$_2$. Previous calculations have attributed the presence of satellite features in the M$_{5,4}$-edge XANES spectra of formally Ce$^{4+}$ compounds to interaction of 3$d^9$4$f^1$ and 3$d^9$4$f^2$ configurations in the final state.~\cite{ra92, ani13} In this general sense the M$_{5,4}$-edge XANES spectrum of CeNi$_2$P$_2$ resembles that of the molecular compound (Et$_4$N)$_2$CeCl$_6$ in addition to extended solids and intermetallics such as CeO$_2$ and CeRh$_{3}$.~\cite{kaindl84} Because transitions associated with the 3$d^9$4$f^1$ and 3$d^9$4$f^2$ final states are not well-resolved in the M$_{5,4}$-edge spectra, the intensity of the satellite features cannot be directly related to the amounts of Ce$^{4+}$, or 4$f^0$ character in the ground state for CeNi$_2$P$_2$. However, the presence of small satellite features in the Ce M$_{5,4}$-edge spectrum for CeNi$_2$P$_2$, and the lack thereof for CePd$_2$P$_2$, is consistent with the observation of $<$ 10\% Ce$^{4+}$ character in the ground state of CeNi$_2$P$_2$.

\begin{figure}[!tht]
	\begin{center}
		\includegraphics[width=1\linewidth]{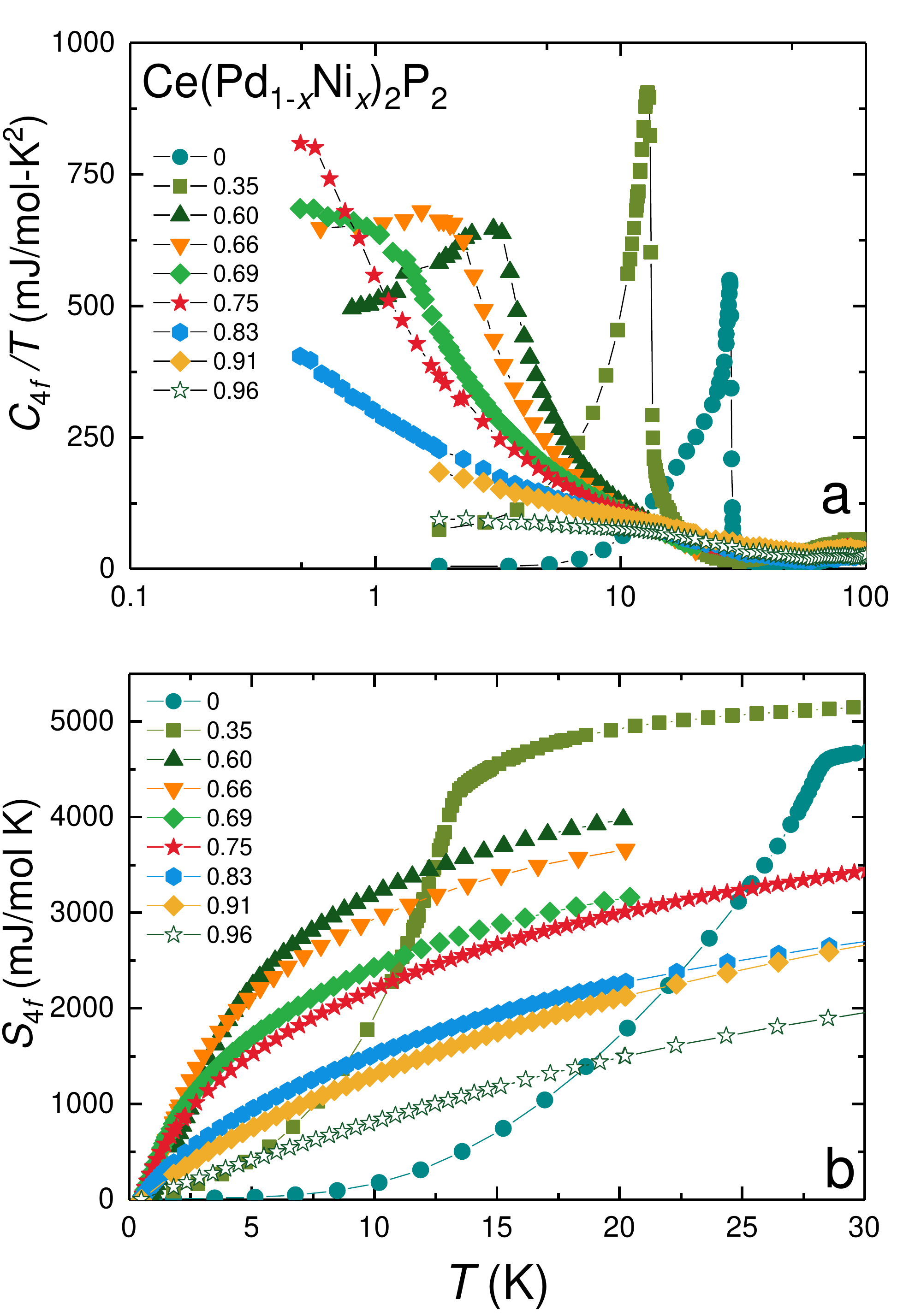}
		\caption{ (a) The heat capacity $C_{4f}$ divided by temperature $T$ vs $T$ following the phonon background subtraction for select concentrations of Ce(Pd$_{1-x}$Ni$_x$)$_2$P$_2$, $C_{4f}/T$ is calculated by subtracting $C/T$ for the nonmagnetic analogue La(Pd$_{1-x}$Ni$_x$)$_2$P$_2$ from that of Ce(Pd$_{1-x}$Ni$_x$)$_2$P$_2$. (b) 4$f$ entropy $S_{\rm{4f}}$ vs $x$. $S_{\rm{4f}}$ is obtained from the heat capacity data as described in the text. } 
		\label{HC}
	\end{center}
\end{figure}

The heat capacity $C_{4f}$ divided by $T$ vs. $T$ data are shown in  Fig.~\ref{HC}a, which further expose the ordered state and underlying electronic behavior. The $x$ $=$ 0 ferromagnetism appears as a lambda-like feature near $T_{\rm{C}}$ $=$ 28.5 K, consistent with a second order phase transition. With increasing $x$, $T_{\rm{C}}$ moves to lower temperatures and up to $x$ $\approx$ 0.35 the shape of phase transition is preserved but its overall size grows. This indicates that even as $T_{\rm{C}}$ is suppressed the associated entropy is conserved. Between 0.35 $<$ $x$ $\lesssim$ 0.69, the ferromagnetic feature broadens and is superimposed on an increasing background. The broadening of the phase transition is attributed to chemical/structural disorder which is maximal near the middle of the substitution series. As $T_{\rm{C}}$ approaches zero near $x_{\rm{cr}}$ $\approx$ 0.7, $C_{4f}/T$ diverges nearly continuously down to 0.5 K. This is a common feature of `non-Fermi-liquid' behavior near a quantum critical point in correlated $f$-electron materials and may be associated with quantum critical fluctuation of the magnetic order parameter.~\cite{Stewart01,Rosch07,Gegenwart08,Pfleiderer09,brando16} For larger $x$ the divergence weakens and finally tends to saturate at low temperature for $x$ = 0.96 in a manner that is similar to  CeNi$_2$P$_2$, indicating the recovery of the paramagnetic Fermi liquid state.  

The 4$f$ contribution to the entropy $S_{4f}$ vs. $T$ is shown in Fig.~\ref{HC}b. $S_{4f}$ was calculated by subtracting $C/T$ for the nonmagnetic analogue La(Pd$_{1-x}$Ni$_x$)$_2$P$_2$ from that of Ce(Pd$_{1-x}$Ni$_x$)$_2$P$_2$ and subsequently integrating from 0.5 K. The nonmagnetic lattice term was approximated by summing the heat capacities of LaPd$_2$P$_2$ and LaNi$_2$P$_2$ in the ratios (1-$x$):$x$. While this approach slightly underestimates the total 4$f$ entropy and only approximates the lattice contribution to the heat capacity, it provides a consistent way to assess the evolution of $S_{4f}$ with $x$. $S_{4f}$ reaches 0.88$R$ln2 at $T_{\rm{C}}$ for $x$ $=$ 0. This is slightly reduced from the full entropy of a a doublet ground state and indicates weak Kondo screening of the $f$-moment by the conduction electrons.~\cite{tran14,ikeda15} In the $x$-region where the phase transition remains sharp (0 $\leq$ $x$ $\lesssim$ 0.35), $S_{4f}$ consistently recovers to similar values at $T_{\rm{C}}$, suggesting that the strength of the hybridization changes little over this range. For specimens with larger concentrations that still show ferromagnetism but have broadened phase transitions (0.35 $\lesssim$ $x$ $\lesssim$ 0.69), the entropy recovered at $T_{\rm{C}}$ grows smaller with increasing $x$, revealing strengthening hybridization. For concentrations in the no order region ($x$ $\gtrsim$ $x_{\rm{cr}}$) $S_{4f}$ is significantly reduced from that seen at lower $x$ and increases smoothly with increasing $T$ in a manner consistent with there being strong Kondo hybridization between the $f$- and conduction electrons.\cite{Doniach_77,kondo}

\begin{figure}[!tht]
    \begin{center}
        \includegraphics[width=3.5in]{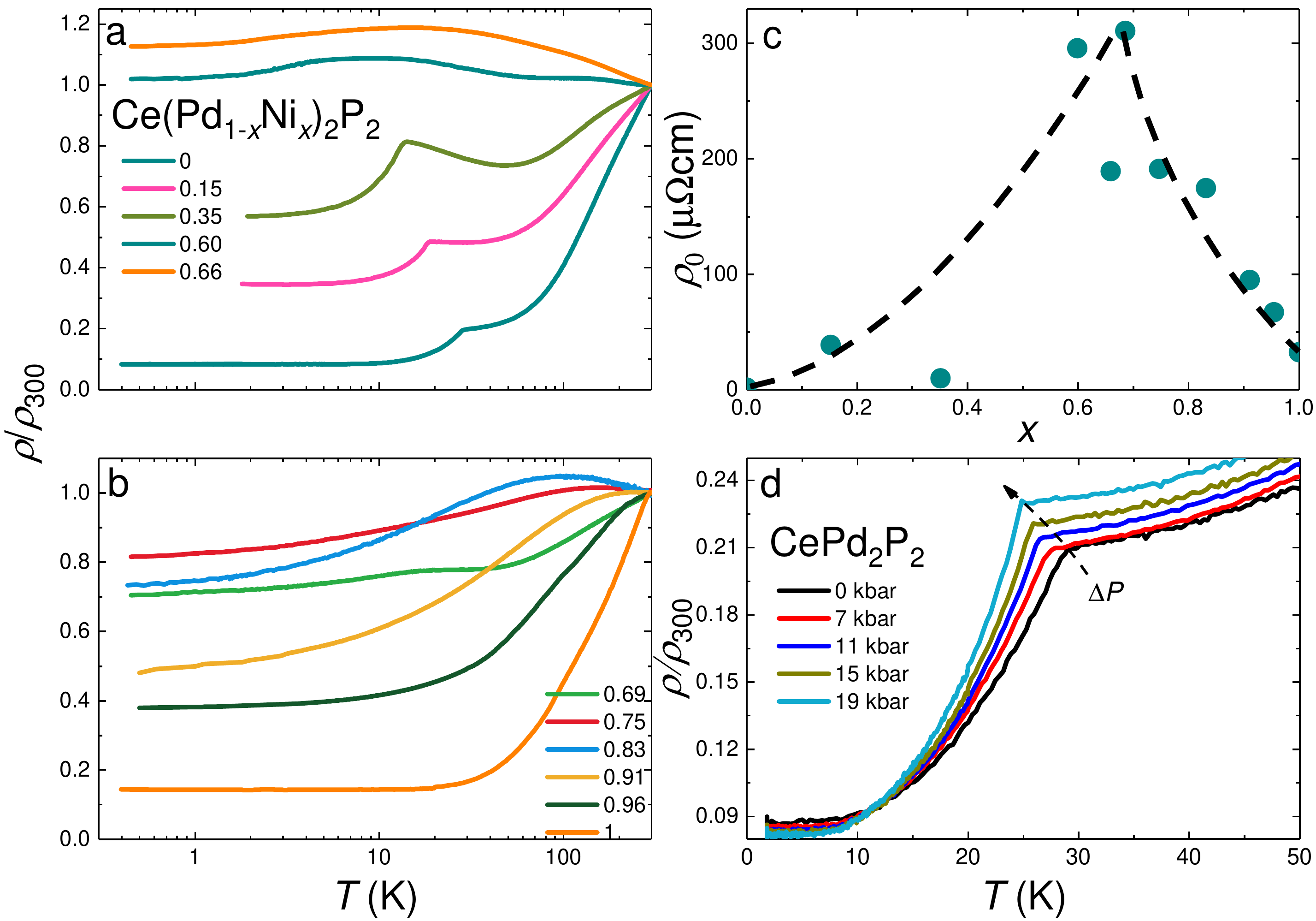}
        \caption{(a) and (b) The electrical resistivity normalized to the room temperature value $\rho/\rho_{\rm{300K}}$ vs. temperature $T$ for Ce(Pd$_{1-x}$Ni$_x$)$_2$P$_2$ at values $x$ $=$ 0 $-$ 1. (c) The residual resistivity $\rho_0$ vs $x$. (d) $\rho/\rho_{\rm{300K}}$($T$) collected under applied pressure $P$ $\lesssim$ 19 kbar for CePd$_2$P$_2$.
        }
        \label{rho}
    \end{center}
\end{figure}

The temperature dependences of the electrical resistivity normalized to the room temperature value $\rho/\rho_{300K}$ vs. $T$ for the entire substitution series are shown in Fig.~\ref{rho}. The behavior for $x$ $=$ 0 is consistent with earlier results, where the resistivity decreases with decreasing $T$ and evolves through a kink near $T_{\rm{C}}$ $=$ 28.5 K that further reduces the electronic scattering due to the removal of magnetic fluctuations.~\cite{tran14} Here, the residual resistivity ratio $RRR$ $=$ $\rho/\rho_{\rm{300K}}$ $\approx$ 12. For 0 $\leq$ $x$ $\lesssim$ 0.35 $RRR$ decreases due to increasing disorder, but the reduction in $\rho$/$\rho_{300K}$ at $T_{\rm{C}}$ remains sharp. The $x$ - dependence of the residual resistivity $\rho_0$ is shown in Fig.~\ref{rho}c, where the doping introduces a substantial amount of disorder which results in a large residual resistivity near the critical region ($\rho_0$ $\approx$ 300 \si{\micro \ohm}cm). Based on this, we estimate that the specimens in this concentration range belong to the second regime as described in the BKV theory,~\cite{brando16, Belitz12} where $\rho_0$ is several hundred \si{\micro \ohm}cm. Over this $x$-range the phase transition is preceded in temperature a growing upturn in $\rho$/$\rho_{300K}$ which indicates a gradual strengthening hybridization between the $f$- and conduction electron states. For larger $x$ the phase transition broadens due to increasing disorder and continues to be suppressed until it is no longer visible near $x_{\rm{cr}}$ $\approx$ 0.7.

In order to further examine the tuning mechanisms that control the ordered state in CePd$_2$P$_2$, we performed measurements of the electrical resistivity under hydrostatic pressure (Fig.~\ref{rho}d). We find that the ferromagnetic phase transition is monotonically suppressed with increasing pressure at a rate of 2.4 K/GPa, and from this we estimate that an applied pressure of 12 GPa would be needed to fully suppress the ferromagnetism to zero temperature in the parent compound. In order to directly compare this result to what is seen for Pd $\rightarrow$ Ni substitution, we convert the applied pressure to change in unit cell volume and then associate this value with a Ni concentration. Results for a typical bulk modulus $B_0$ $\approx$ 110 GPa (similar to what is observed for CeCu$_2$Si$_2$~\cite{spain86}) are shown as open stars in Fig.~\ref{phase}a, where the slope of $T_{\rm{C}}$ is weaker than that seen in the substitution series.

\begin{figure}[!tht]
    \begin{center}
        \includegraphics[width=3.5in]{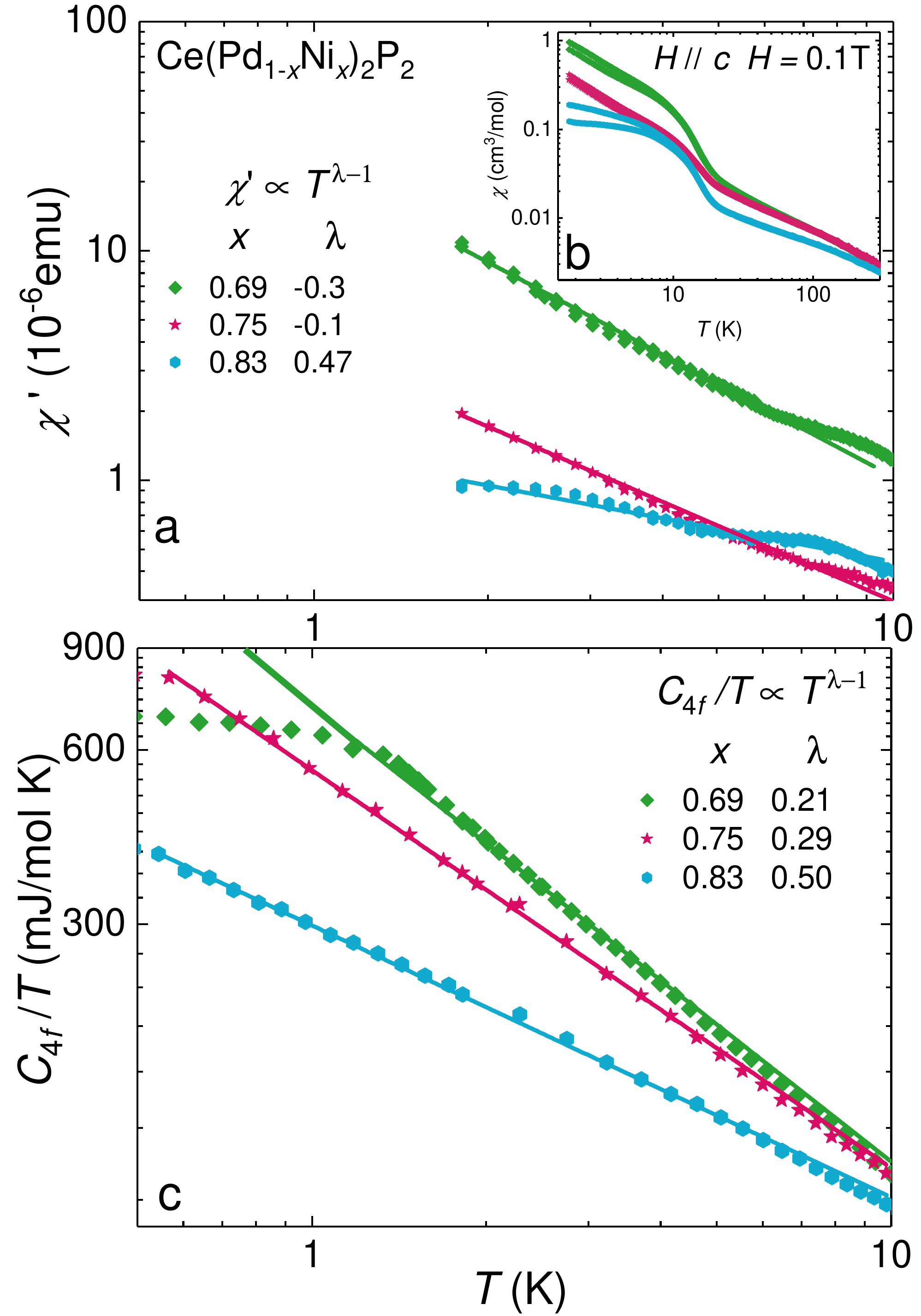}
       \caption{(a) AC susceptibility $\chi$' of selected selected $x$ of Ce(Pd$_{1-x}$Ni$_x$)$_2$P$_2$ series. Solid lines are fits to the data using the expression $\chi$' $\propto$ $T^{\lambda-1}$. (b) Magnetic susceptibility $\chi$ with $H$ $=$ 0.1 T applied parallel $\parallel$ $c$ of selected selected $x$ of Ce(Pd$_{1-x}$Ni$_x$)$_2$P$_2$ series. (c) $C_{4f}/T$ for selected $x$ at low temperatures. Solid lines are fits to the data using the expression $C_{4f}/T$ $\propto$ $T^{\lambda-1}$. 
       }
        \label{griffith}
    \end{center}
\end{figure}

Finally, in order to assess whether chemical disorder produces magnetic clustering behavior that impacts the low temperature behavior in the large $x$-region, AC magnetic susceptibility $\chi$' measurements were performed for selected concentrations (Fig~\ref{griffith}a). For $x$ $>$ $x_{\rm{cr}}$ the data can be fit using the formula $\chi$' $\propto$ $T^{\lambda-1}$, where $\lambda$ becomes less negative with increasing $x$ and changes sign to become positive for $x$ $=$ 0.83. This type of behavior is expected if chemical disorder produces cluster regions with short range magnetic correlations while the bulk state remains paramagnetic: e.g., as for a quantum Griffiths phase (QGP).~\cite{Griffiths1969,vojta,sereni07, Westerkamp09} Similar fits were carried out for $C/T$ over a broader temperature range, which also reveal a systematic evolution in $\lambda$ that might be consistent with a QGP. Over the same $x$-range, we find that there is a weak and hysteretic increase in $\chi$($T$) that disappears before $x$ = 1 (Fig. 7b). While this feature indicates the persistence of short range ferromagnetic interactions for $x$ $>$ $x_{\rm{cr}}$, it does not appear in other bulk measurements such as the heat capacity. We furthermore point out that the power-law behavior extends over a broad $x$-range, which is in contrast to the contained v-shaped region that is often seen for ordered materials with quantum critical behavior~\cite{Stewart01,Rosch07,Gegenwart08,Pfleiderer09}. While these measurements are suggestive of QGP behavior, further work is still needed: e.g., the trends at lower temperatures should be established.

\begin{figure}[!tht]
    \begin{center}
        \includegraphics[width=3.5in]{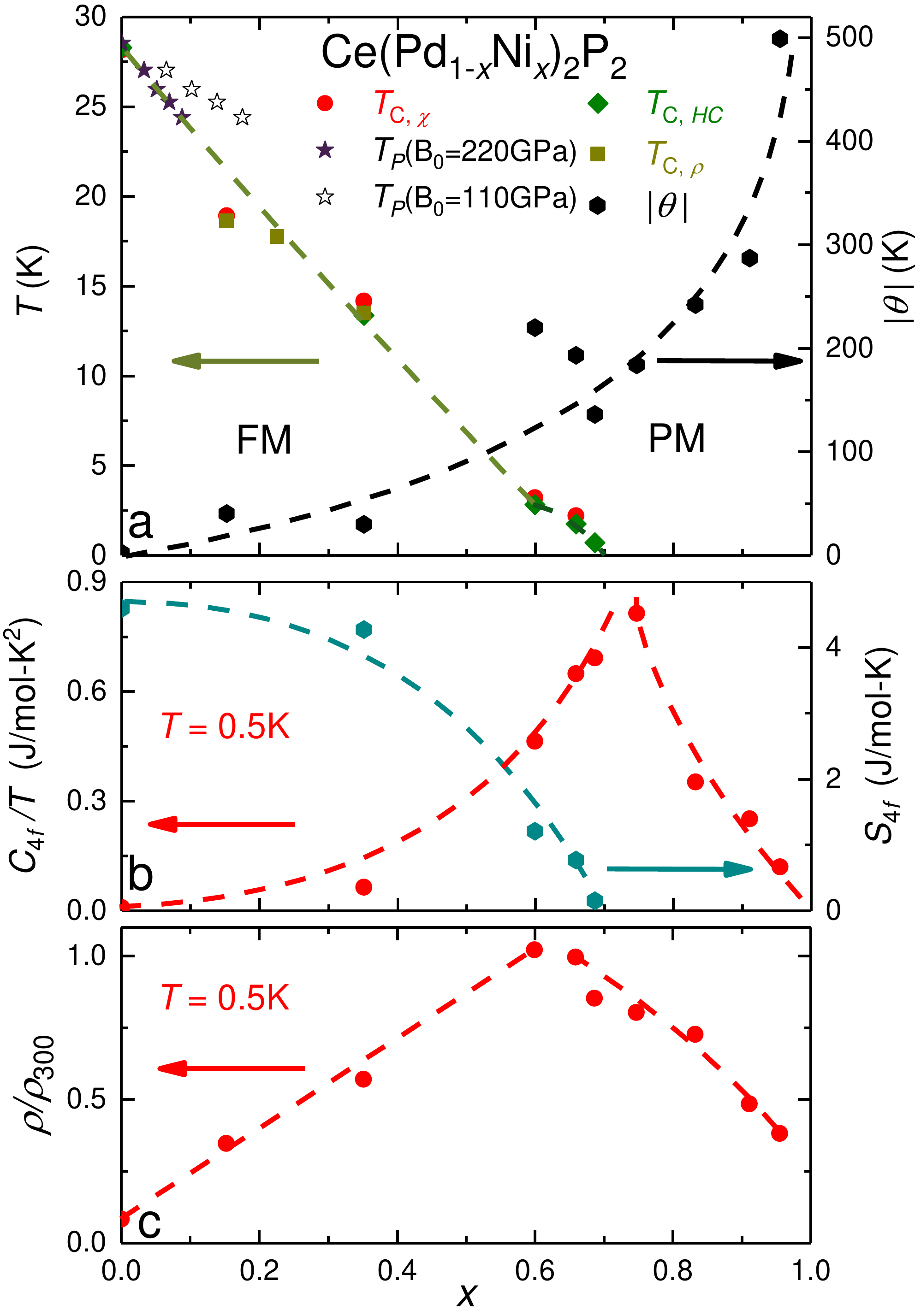}
        \caption{(a) left axis: Temperature $T$ vs. concentration $x$ phase diagram for Ce(Pd$_{1-x}$Ni$_x$)$_2$P$_2$ for $x$ $=$ 0 $-$ 1 constructed from magnetic susceptibility $\chi$ $=$ $M/H$, electrical resistivity $\rho$ and heat capacity $C$ data. Also shown is the ferromagnetic phase boundary that was observed for CePd$_2$P$_2$ under applied pressure. The Birch-Murnaghan equation of state was used to convert from pressure to unit cell volume, and then to the corresponding $x$. The solid and open stars are for $B_0$ $=$ 110 and 220 GPa, respectively. Right axis: Averaged Curie-Weiss temperature $|$ $\theta$ $|$ vs $x$. (b) Left axis: The value of the background subtracted heat capacity divided by temperature $C_{4f}/T$ at $T$ $=$ 0.5 K. Right axis: The 4$f$ contribution to the entropy at $T_{\rm{C}}$ vs $x$. (c) Electrical resistivity normalized to the room temperature value $\rho$/$\rho_{\rm{300K}}$ at $T$ $=$ 0.5 K vs $x$.}
        \label{phase}
    \end{center}
\end{figure}

\section{\label{sec:level1}Discussion}
Fig.~\ref{phase} shows the $T$-$x$ phase diagram and the evolution of several quantities vs. $x$. $T_{\rm{C}}$ is suppressed linearly with $x$, and is extrapolated to approach zero temperature near $x_{\rm{cr}}$ $\approx$ 0.7. For $x$ $=$ 0 we also plot results from measurements under applied pressure, where chemical and applied pressure both continuously suppress $T_{\rm{C}}$. If we assume $B_0$ $=$ 220 GPa (closed stars) then both chemical and applied pressure suppress $T_{\rm{C}}$ at the same rate. A more realistic value of $B_0$ $=$ 110 GPa (open stars) results in a more gradual suppression of $T_{\rm{C}}$. Regardless of which $B_0$ is chosen, it is clear that the main tuning parameter that controls $T_{\rm{C}}$ is the unit cell volume, which likely changes the relative strengths of the Kondo and RKKY interactions in a Doniach-like scenario.\cite{Doniach_77} This argument is strengthened by considering that for Kondo lattice systems an estimate of the Kondo energy scale can be made using the expression $\chi(T)$ = $C/(T-2T_K)$.~\cite{thy75} As shown Fig.~\ref{phase}a, $\theta$ increases with increasing $x$, suggesting a strengthening Kondo energy scale. For $x$ $\gtrsim$ 0.7 there is also evidence that the $f$-electron state is distinct from what is seen for $x$ $<$ $x_{\rm{cr}}$: i.e., the cerium $f$-valence evolves away from a purely trivalent state. This is revealed through: (1) a deviation from Vegard's law and (2) in the XANES measurements of the end-member compounds, which show that CePd$_2$P$_2$ has a 3+ $f$-electron valence, while CeNi$_2$P$_2$ shows an admixture of 3+ and 4+. A more detailed study of the region near $x_{\rm{cr}}$ is still needed to determine whether or not this change in the substitution series is abrupt or gradual.

An important feature of this substitution series is that the phase transition remains second order across the entire ferromagnetic $x$-regime. This is likely because disorder influences the intermediate substitution region: e.g., as seen in the broadening of the phase transition in heat capacity and the growing residual resistivity that peaks near $x$ $\approx$ 0.6. This provides the conditions that are expected from the BKV theory for a disordered ferromagnetic quantum critical point near $x_{\rm{cr}}$ $\approx$ 0.7. At the same time, there is some evidence that the disorder leads to magnetic clustering behavior that might be consistent with a quantum Griffiths phase. Further work is needed to verify this scenario and its impact on the possible quantum critical behavior, such as measurements at lower temperatures to establish the powerlaw behavior. In future work, it will also be useful to compare to clean and lightly substituted CePd$_2$P$_2$ under applied pressure, where the evolution from first order to second order behavior can be systematically followed.

Finally, we make a comparison to some related materials. For instance, the prototypical antiferromagnetic quantum critical point alloy series CeCu$_2$Si$_{2-x}$Ge$_x$,~\cite{Knebel} shows a qualitatively similar evolution of the low temperature phenomena. In particular, $C/T$ for $x_{\rm{cr}}$ follows a logarithmic in temperature divergence with a value near 0.9 J/mol K$^2$ at low temperatures that is replaced by antiferromagnetic order with increasing unit cell volume. An important difference is that for this system the critical region is near $x$ $=$ 0 and the amount of disorder is small by comparison to what is seen in our series. Another closely related alloy series is CePd$_2$As$_{2-x}$P$_x$,~\cite{shang} which features a transformation from ferromagnetic to antiferromagnetic order going from P $\rightarrow$ As but does not have obvious features associated with quantum criticality or a quantum Griffiths phase. In this series, it appears that the nonisoelectronic chemical substitution mainly tunes the sign of the magnetic exchange. We finally point out that although the alloy series CeRh$_{1-x}$Pd$_x$~\cite{sereni07} crystallizes in a different structure, there are remarkable similarities to what we have observed for Ce(Pd$_{1-x}$Ni$_x$)$_2$P$_2$: including that the ferromagnetic order is suppressed towards zero temperature in the disordered $x$-region. In the same area of the phase diagram there is evidence for a change of the cerium valence and the presence of a quantum Griffiths phase. It will be useful to study these systems together to assess the universality of model for ferromagnetic quantum criticality, including the BKV theory.

\section{\label{sec:level1}Conclusions}
These results reveal that Ce(Pd$_{1-x}$Ni$_x$)$_2$P$_2$ is a useful example of a cerium-based intermetallic with a disordered ferromagnetic QCP and accompanying breakdown of Fermi liquid behavior. We also find that for CePd$_2$P$_2$ a pressure of $P_{\rm{c}}$ $\approx$ 12 GPa would likely be sufficient to access the tricritical point and first order quantum phase transition that is expected in the clean limit. It will be interesting to compare the electronic states that appear near $x_{\rm{cr}}$ and $P_{\rm{cr}}$ to test expectations from BKV theory. Further comparison to more conventional antiferromagnetic QCPs is also of interest, where an important question is whether unconventional superconductivity can occur near a disordered ferromagnetic QCP.~\cite{belitz99, belitz05, belitz15, brando16,huang13, Nakatsuji08,Griffiths1969,vojta,sereni07, Westerkamp09}

\vspace{3mm}
\section{\label{sec:level1}Acknowledgements}
A portion of this work was performed at the National High Magnetic Field Laboratory (NHMFL), which is supported by National Science Foundation Cooperative Agreement No. DMR-1157490 and DMR-1644779, and the State of Florida. Research of RB, YL, DG, KH, WP, WLN, SAK and TAS were supported in part by the Center for Actinide Science and Technology, an Energy Frontier Research Center funded by the U.S. Department of Energy (DOE), Office of Science, Basic Energy Sciences (BES), under Award Number DE-SC0016568. MC was supported by DOE-BES through Award No. DE-339SC0002613. Portions of this work were supported by the LANL named fellowship program; the Agnew National Security Fellowship (SEB) and the Glenn T. Seaborg Institutes Postdoctoral Fellowship program at LANL (MGF). Use of the Stanford Synchrotron Radiation Lightsource, SLAC National Accelerator Laboratory, was supported by the U.S. Department of Energy, Office of Science, Office of Basic Energy Sciences under Contract No. DE-AC02-76SF00515. The SSRL Structural Molecular Biology Program is supported by the DOE Office of Biological and Environmental Research, and by the National Institutes of Health, National Institute of General Medical Sciences (including P41GM103393). This work was also supported by the Joint Plasma Physics Program of the National Science Foundation and the Department of Energy under grant DE-SC0016251 (GTS). SM was supported by the Director, Office of Science, Office of Basic Energy Sciences, Division of Chemical Sciences, Geosciences, and Biosciences Heavy Element Chemistry Program of the U.S. Department of Energy (DOE) at LBNL under Contract No. DE-AC02-05CH11231. M$_{5,4}$-edge spectra described in this paper was measured at the Canadian Light Source, which is supported by the Canada Foundation for Innovation, Natural Sciences and Engineering Research Council of Canada, the University of Saskatchewan, the Government of Saskatchewan, Western Economic Diversification Canada, the National Research Council Canada, and the Canadian Institutes of Health Research.

\end{document}